\documentclass[oldversion]{aa}  
\usepackage{graphicx,psfig,amsmath}

\begin{document}

   \title{Evaporation of the planet HD\,189733b observed in H\,{\sc i} Lyman-$\alpha$}

   \author{
   A.~Lecavelier des Etangs\inst{1,2}
 \and
D. Ehrenreich\inst{3}
\and
   A.~Vidal-Madjar\inst{1,2}       
\and
G.~E.~Ballester\inst{4}
\and
J.-M.~D\'esert\inst{1,2}       
\and
R.~Ferlet\inst{1,2}       
\and
G.~H\'ebrard\inst{1,2}       
\and
D.~K.~Sing\inst{1,2,5}      
\and
K.-O.~Tchakoumegni\inst{1,2,6}
\and
S.~Udry\inst{7}        
 }
   

\offprints{A.L. (\email{lecaveli@iap.fr})}

   \institute{
   CNRS, UMR 7095, 
   Institut d'astrophysique de Paris, 
   98$^{\rm bis}$ boulevard Arago, F-75014 Paris, France
   \and
   UPMC Univ. Paris 6, UMR 7095, 
   Institut d'Astrophysique de Paris, 
   98$^{\rm bis}$ boulevard Arago, F-75014 Paris, France
   \and
   Universit\'e Joseph Fourier - Grenoble 1 / CNRS,  
   Laboratoire d'astrophysique de Grenoble (LAOG) UMR 5571, 
   BP 53, 38041 Grenoble cedex 09, France 
   \and
   Lunar and Planetary Laboratory, University of Arizona, 1541 E. University Blvd., Tucson, AZ 85721-0063, USA
   \and
Astrophysics Group, School of Physics, University of Exeter, Stocker Road, Exeter EX4 4QL, UK
   \and
   Lyc\'ee Passy Buzenval, 50 avenue Otis Mygatt, 92508 Rueil-Malmaison Cedex, France
    \and
   Observatoire de Gen\`eve, Universit\'e de Gen\`eve, 51 chemin des Maillettes, 1290 Sauverny, Switzerland
   }
   
   \date{} 
 
  \abstract
{
We observed three transits of the extrasolar planet HD\,189733b in H\,{\sc i} Lyman-$\alpha$ and in a few other lines in the ultraviolet with HST/ACS, in the search for atmospheric signatures. 
We detect a transit signature in the Lyman-$\alpha$ light curve 
with a transit depth of 5.05$\pm$0.75\%. 
This depth exceeds the occultation depth produced by the planetary disk alone at the 3.5$\sigma$ level (statistical).
Other stellar emission lines are less bright, and, taken individually, they do not show the transit signature, while the whole spectra redward of the Lyman-$\alpha$ line has enough photons to show a transit signature consistent with the absorption by the planetary disk alone.
The transit depth's upper limits in the emission lines are 11.1\% for O\,{\sc i} $\lambda$1305\AA\ and 5.5\% for C\,{\sc ii} $\lambda$1335\AA\ lines.

The presence of an extended exosphere of atomic hydrogen around HD\,189733b producing 5\% absorption 
of the full unresolved Lyman-$\alpha$ line flux shows that the
planet is losing gas. The Lyman-$\alpha$ light curve 
is well-fitted by a numerical simulation 
of escaping hydrogen in which the planetary atoms are pushed 
by the stellar radiation pressure. 
We constrain the escape rate of atomic hydrogen to be between 
$10^9$ and $10^{11}$\,g\,s$^{-1}$ and the ionizing extreme UV flux 
between 2 and 40 times the solar value (1-$\sigma$), 
with larger escape rates corresponding to larger EUV flux. 
The best fit is obtained for $\mathrm{d}M/\mathrm{d}t$=$10^{10}$\,g\,s$^{-1}$ and
an EUV flux $F_{\rm EUV}$=20 times the solar value. 
HD\,189733b is the second extrasolar planet for which atmospheric 
evaporation has been detected. 
%
}


   \maketitle
%

\section{Introduction}
 \label{Introduction}

\subsection{Evaporation of hot Jupiters}

Observations of the extrasolar planet HD209458b in the Lyman-$\alpha$ line 
of atomic hydrogen (H\,{\sc i}) 
have revealed that this planet is losing gas (Vidal-Madjar et al.\ 2003). 
Indeed, since the discovery of 51\,Peg\,b (Mayor \& Queloz 1995),
the issue of the evaporation of hot Jupiters (also named ``Pegasides'', Guillot et al.\ 1996) 
has been raised by Burrows \& Lunine (1995) and Guillot et al.\ (1996).
Nonetheless, the discovery of a large number of massive
hot Jupiters\footnote{Usually defined as massive planets orbiting
in less than 10~days} and
very hot Jupiters\footnote{Usually defined as planets orbiting in less than 3~days
({\it e.g.},  Gaudi et al.\ 2005)}
has led to the conclusion that the evaporation {of massive planets} has to be modest.

In this frame the discovery that the transiting extrasolar planet HD\,209458b is indeed losing
mass was rather unexpected.
Lyman-$\alpha$ observations of the planetary transit showed excess absorption
due to an extended H\,{\sc i} cloud and provided a lower limit of the H\,{\sc i} escape rate
on the order of $10^{10}$\,g\,s$^{-1}$ (Vidal-Madjar et al.\ 2003;
Vidal-Madjar \& Lecavelier des Etangs 2004).
This discovery has been challenged by Ben-Jaffel (2007), 
but the apparent discrepancy has been resolved and the result obtained 
by Ben-Jaffel on this first data set 
strengthens the evaporation scenario (Vidal-Madjar et al.\ 2008).
Two subsequent observations at low spectral resolution 
confirm the evaporation with the detection of a 5\% absorption
of the whole Lyman-$\alpha$ line, as measured using unresolved emission line flux during
planetary transits observed with the Hubble Space Telescope (HST) STIS 
(Vidal-Madjar et al.\ 2004) and ACS instruments (Ehrenreich et al.\ 2008).
An independent, new analysis of the low-resolution data set 
used by Vidal-Madjar et al.\ (2004) has confirmed that the transit depth 
in Lyman-$\alpha$ is significantly greater than the transit depth due to 
the planetary disk alone (Ben-Jaffel, private communication). 
All of these three independent observations show 
a significant amount of gas at velocities exceeding the planet escape velocity,
leading to the conclusion that HD\,209458b is evaporating.

These observational constraints have been used to develop 
several theoretical models to explain 
and characterize the evaporation processes (Lammer et al. 2003;
Lecavelier des Etangs et al. 2004, 2007, 2008a; 
Baraffe et al. 2004, 2005, 2006; Yelle 2004, 2006; 
Jaritz et al. 2005; Tian et al. 2005;  
Garc\'ia-Mu\~noz 2007; Holmstr\"om  et al.\ 2008; 
Stone \& Proga 2009; Murray-Clay et al.\ 2009). 
Except for the model of Holmstr\"om  et al.\ (2008) 
in which the observed hydrogen originates from charge
exchange between a stellar wind of exotic properties 
and the planetary escaping exosphere 
(in this model the escape is found to be of the same order of magnitude), 
all of these modeling efforts have led to the
conclusion that most of the EUV and X-ray energy input by
the host star is used by the atmosphere to escape the planetary
gravitational potential. Therefore, an ``energy diagram'' has been developed
to describe the evaporation status of extrasolar planets,
a diagram in which the potential energy of the planet is plotted versus the
energy received by its upper atmosphere (Lecavelier des Etangs 2007).
It has also been suggested that, in the case of planets
with a mass of only a fraction of the mass of a hot Jupiter
and orbiting close to their parent stars,
evaporation can lead to significant modification of the nature of the planet,
leading to the formation of planetary remnants (Lecavelier des Etangs et al.\ 2004).
The recently discovered CoRoT-7b (L\'eger et al.\ 2009) could 
belong to a new category of planets, which was proposed for classification 
as ``chthonian planets'' (see Lecavelier des Etangs et al.\ 2004).
A review of the evaporation of hot Jupiters can be found in Ehrenreich (2008).

However, up to now HD\,209458b has remained the
only planet for which an evaporation process has been observed.
This raises the question of the
evaporation state of other hot Jupiters and very hot Jupiters.
Is the evaporation specific to HD\,209458b or general to hot Jupiters?
How does the escape rate depend on the planetary system and stellar characteristics?
The discovery of HD\,189733b, a planet transiting a bright and nearby K0 star (V=7.7), offers the
unprecedented opportunity to answer these questions. Indeed, among the stars harboring transiting planets,
HD\,189733 belongs to the brightest at Lyman-$\alpha$; therefore, 
despite the failure of the HST/STIS instrument one year before the discovery of HD\,189733b, 
this planet remained one of the best targets when searching for atmospheric escape.
In this paper we present transit observations of HD\,189733b made with the UV channel of 
the Advanced Camera for Surveys onboard the Hubble Space Telescope (HST/ACS) using
the same methodology as employed by Ehrenreich et al.\ (2008) for HD\,209458b.

\subsection{HD\,189733b}

Located just 19.3~parsecs away with a semi-major axis 
of 0.03~AU and an orbital period of 2.2 days, HD\,189733b belongs to the class of 
``very hot Jupiters" (Bouchy et al.\ 2005). 
More important, since this planet is seen to
transit its parent star, the planetary transits and eclipses can be
used to probe the planet's atmosphere and environment 
({\it e.g.}, Charbonneau et al.\ 2008; D\'esert et al.\ 2009).    

HD\,189733b orbits a bright main-sequence star and shows a 
transit occultation depth of $\approx$2.4\% at optical wavelengths
(Pont et al.\ 2007). The
planet has a mass $M_p$=1.13~Jupiter masses ($M_{\rm Jup}$) and a
radius $R_p$=1.16~Jupiter radii ($R_{\rm Jup}$) in the visible (Bakos
et al.\ 2006; Winn et al.\ 2007). The short period of the planet 
(2.21858~days) has been measured precisely (H\'ebrard \& Lecavelier des 
Etangs 2006; Knutson et al.\ 2009). 
Spectropolarimetry has measured the strength and topology 
of the {\it stellar} magnetic field, which reaches up to 40~G (Moutou et al.\ 2007).
Sodium has been detected in the planet's atmosphere
by ground-based observations (Redfield et al.\ 2008). 
Using HST/ACS, Pont et al.\ (2008) detected atmospheric haze, 
which is interpreted as Mie
scattering by small particles (Lecavelier des Etangs et al.\ 2008b). 
High signal-to-noise HST/NICMOS observations (Sing et al.\ 2009) 
have recently shown that the near-IR spectrum below 2~$\mu$m is
dominated by haze scattering rather than by water absorption (Swain et al.\ 2008).
It has been tentatively proposed that CO molecules can explain the excess 
absorption seen at 4.5~$\mu$m (Charbonneau et al.\ 2008; D\'esert et al.\ 2009). 
Using Spitzer spectroscopy of planetary eclipses, the infrared spectra 
of HD\,189733b's atmosphere have revealed signatures of H$_2$O absorption 
and possibly weather-like variations in the atmospheric conditions 
(Grillmair et al.\ 2008). A sensitive search with GMRT has 
provided very low upper limits on the meter-wavelength radio emission from the planet,
indicating a weak planetary magnetic field (Lecavelier des Etangs et al.\ 2009). 

\section{Observations}

\begin{table*}[tbh]
\begin{tabular}{clcccc}
\hline
\hline
\noalign{\smallskip}
Data set & Date & \multicolumn{2}{c}{Observation} & \multicolumn{2}{c}{Transit}  \\
         &      & Start & End & Start & End  \\
\noalign{\smallskip}
\hline
\noalign{\smallskip}

Transit \#1 & 2007-06-10     & 02:04:52 & 07:35:03 & 03:36:59 & 05:24:28 \\
Transit \#2 & 2007-06-18/19  & 22:43:24 & 04:12:45 & 00:35:58 & 02:23:27 \\
Transit \#3 & 2008-04-24     & 14:01:12 & 17:53:22 & 15:00:29 & 16:47:58 \\
\noalign{\smallskip}
\hline
\hline
\end{tabular}
\caption{Log of the observations. Time is given in UT.}
\label{Obs Log}
\end{table*}

Using the slitless prism spectroscopy mode of the Solar Blind Camera (SBC) of the 
HST/ACS, we observed 3~transits of HD\,189733b. 
These observations provide 2D spectral images covering the far-UV 
wavelengths from $\sim$1200\AA\ to $\sim$1700\AA .
The log of the observations is given in Table~\ref{Obs Log}. 
We observed during four HST orbits for the first two transits, 
and during three HST orbits for the last transit. 
For each series of exposures within a given HST orbit, 
we obtained four individual exposures with the target at a first position on the detector, 
and four other exposures at a second position (dithering).
We thus obtained a total of 32~exposures for transits~\#1 and~\#2 
and 24 exposures for transit~\#3.
Each exposure had 257~seconds effective exposure time. 

\section{Data analysis}

\begin{figure}[tbh]
\psfig{file=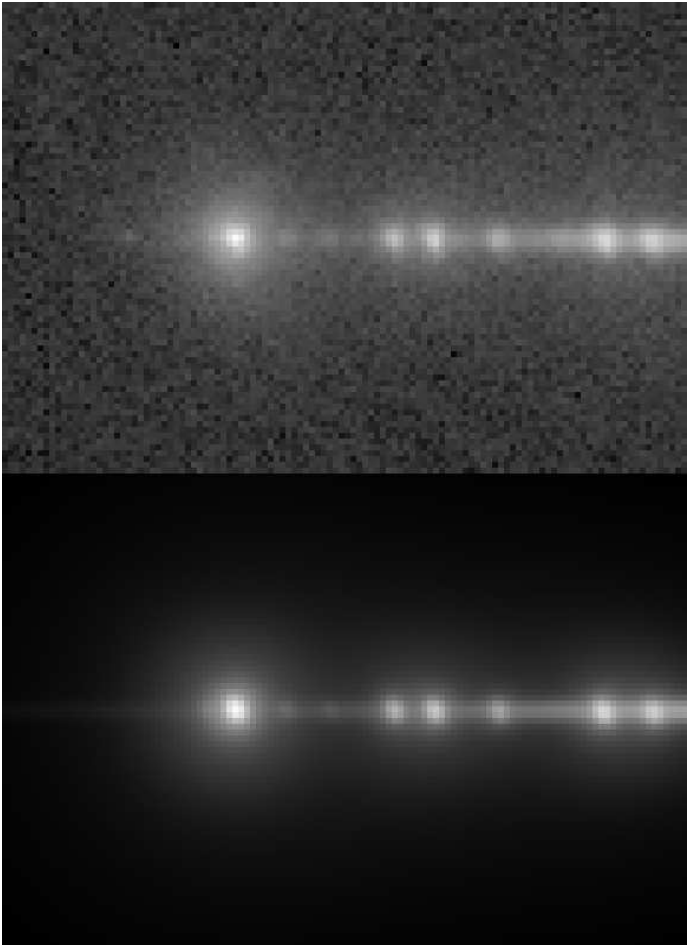,width=\columnwidth} 
\caption[]{2D image of the HD\,189733 spectrum 
obtained with ACS prism spectroscopy (upper panel). 
The bottom panel shows the 2D-model of this image. 
In this image, stellar emission lines can be seen as bright spots: 
from left to right, the six brightest spots are the Lyman-alpha line, 
followed by the two O\,{\sc i} and C\,{\sc ii} emission features, further to the right is the Si\,{\sc iv} feature, 
and the two last spots are emission by C\,{\sc iv} and C\,{\sc i}. 
The stellar continuum emission is seen as a horizontal bright line along the spectrum. 
\label{2D image}}
\end{figure}

\begin{figure}[tbh]
\psfig{file=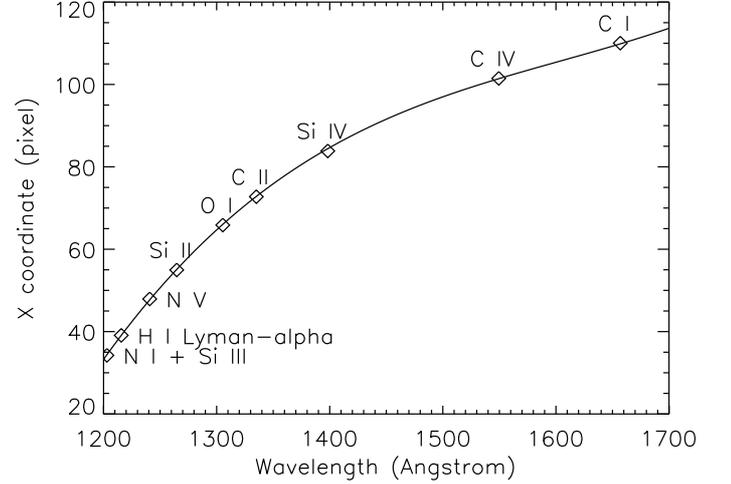,width=\columnwidth,angle=90} 
\caption[]{Plot of the wavelength calibration. The squares show the position of each identified emission feature as a function of wavelength. The solid line shows the fit by a third-degree polynomial. 
\label{Wavelength}}
\end{figure}

\begin{figure}[tbh]
\psfig{file=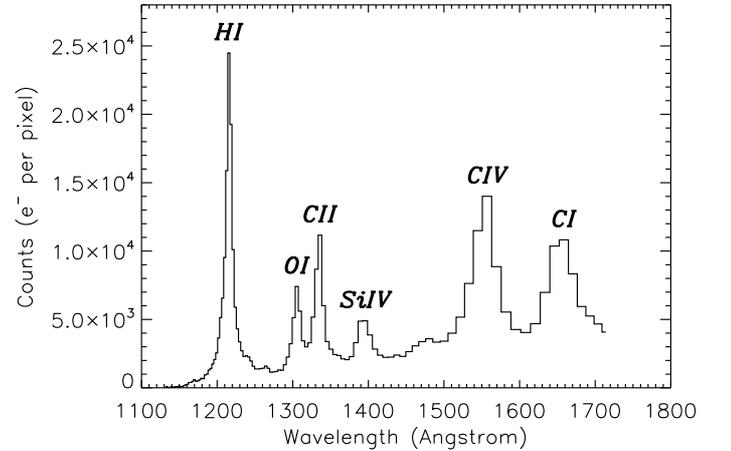,width=\columnwidth,angle=90} 
\caption[]{Plot of one-dimensional wavelength calibrated ACS spectrum of HD\,189733. The brightest lines are identified.
\label{1D plot}}
\end{figure}

The data consist of 2D prism dispersed images of the star HD\,189733. 
At these far-UV wavelengths, the spectrum of a K-type star is composed 
of well-known bright chromospheric emission lines arising 
above a faint photospheric continuum. These emission lines form features 
that are seen as bright spots in the 2D images (Fig.~\ref{2D image}).
The brightest line is the H\,{\sc i} Lyman-$\alpha$ at $\lambda$1216\AA .
By identifying other features (e.g., H\,{\sc i}, O\,{\sc i}, C\,{\sc II}, etc.) 
the spectrum can be wavelength-calibrated. 

The aim of the data analysis is to obtain the best estimate of the brightness 
of each emission feature (with its associated error bar) 
in order to search for a possible transit signature.
The first step is to subtract the background of each image. 
This background is caused mainly by the bright geocoronal Lyman-$\alpha$ airglow that 
produces a roughly uniform background entering the detector in the slitless prism mode. 
This background is accurately determined by the mean flux measured in each image 
above and below the stellar spectrum.
We used two wide zones 100~pixels high centered 100~pixels above and below the spectrum. 
The background level is found to vary significantly within a given HST orbit. 
Nonetheless, various tests show that the background estimate is robust and its associated error
is negligible in the final error budget, well below the photon noise of all the measured
emission features.   

To estimate the brightness of each emission feature, 
we fitted each 2D spectral image with a 
parametric model and calculated the total number of counts in each line after
subtracting the modeled 2D flux from the other emission features and from the continuum.    
For each 2D spectral image, the fit to the image was performed in 3~steps. 
First, we calculated a preliminary point spread function (PSF) 
by fitting the Lyman-$\alpha$ emission line. 
In the second step, the whole image was fitted with a model of six emission features whose 
positions were free to vary. In this fit, the initial guess for the features positions
was set close to the six brightest spots as seen in Fig.~\ref{2D image}
(H\,{\sc i}, O\,{\sc i}, C\,{\sc ii}, Si\,{\sc iv}, C\,{\sc iv}, C\,{\sc i}). 
This second step was used to obtain a preliminary 
wavelength calibration by using the brightest emission lines.
Finally, in a third step, 
we performed a fit of the whole 2D image in which the list of emission lines was extended to fainter lines. 
In this final stage the model included
the wavelengths of the emission lines as a fixed parameter, while the wavelength 
calibration was free to vary and 
was defined by a 3$^{\rm rd}$ order polynomial in the X direction and a 1$^{\rm st}$ order 
polynomial in the Y direction. The line list input to the model 
was defined by trial and error; few misidentifications at the beginning of
the process were easily corrected (even for faint lines) by checking the lines
positions on a wavelength calibration plot. We ended 
up with a line list given in Table~\ref{Line list}.

\begin{table}[tb]
\begin{center}
\begin{tabular}{cl}
\hline
\hline
\noalign{\smallskip}
Wavelength & Identification \\
(\AA ) & \\
\noalign{\smallskip}
\hline
\noalign{\smallskip}
 1203.0   & N\,{\sc i} + Si\,{\sc iii}  \\
 1215.7   & H\,{\sc i} Ly-$\alpha$ \\
 1240.8   & N\,{\sc v} doublet \\
 1264.7   & Si\,{\sc ii}*   \\
 1305.4   & O\,{\sc i}* + O\,{\sc i}** \\ 
 1335.1   & C\,{\sc ii} + C\,{\sc ii}*  \\
 1393.7   & Si\,{\sc iv}  \\
 1402.8   & Si\,{\sc iv}  \\
 1549.5   & C\,{\sc iv} doublet  \\
 1656.9   & C\,{\sc i} multiplet \\
\noalign{\smallskip}
\hline
\hline
\end{tabular}
\end{center}
\caption{List of lines used to fit the 2D images.}
\label{Line list}
\end{table}

The intrinsic width of the observed emission lines 
is much smaller than the prism resolution, so the
lines can be used to estimate the instrumental 
PSF. 
In all three steps of the fitting procedure, 
the PSF was modeled by the addition of six Gaussians. 
The sharpest Gaussian is an asymmetrical 2D Gaussian whose widths
along the two axes and the orientation are free to vary.
This sharp Gaussian has a width of about one pixel, and
was found to vary from one exposure to the next; 
this is likely 
due to the telescope jitter.
The five other Gaussians are axi-symmetrical Gaussians of fixed widths :
$\sigma_1$=2.5\,pixels, $\sigma_2$=4\,pixels, $\sigma_3$=6\,pixels, 
$\sigma_4$=10\,pixels, and $\sigma_5$=25\,pixels.
These widths had been obtained by numerous experiments and allowed us to obtain the best fit to the data. 

The stellar continuum was found to be well-fitted by an exponential law in the form 
$F_{\rm cont} (\lambda)=a \exp (b \lambda)$, where $a$ and $b$ are free parameters.
This stellar continuum model was also convolved with 
the PSF defined in the previous paragraph.  

The resulting wavelength calibration (third degree polynomial) 
is plotted in Fig.~\ref{Wavelength} together with the 
features positions obtained at the second step of the fit. It is clear that the 
3$^{\rm rd}$ degree polynomial gives a satisfactory wavelength calibration. 
This calibration also allow correction for misidentifications. 
For instance, we first believed that the line between the N\,{\sc v} doublet and the O\,{\sc i} emission 
feature was a line of Si\,{\sc ii} from the ground level at $\lambda$1260.4\AA. 
But its measured position was significantly off 
the wavelength calibration. We thus realized that this
line is in fact a line of Si\,{\sc ii}* at $\lambda$1264.7\,\AA , the Si\,{\sc ii} line being 
extincted by the interstellar medium absorption.
To obtain the wavelength calibrated spectrum of HD\,189733 plotted in Fig.~\ref{1D plot}, 
we aligned all 2D images 
using the measured line positions, and
we calculated the total spectrum by adding the measured counts along 
a band of 5~pixels height.

With the PSF parameters (8~parameters), the wavelength calibration (6~parameters), the brightness
of each line (10~parameters for a total of 10~lines), the stellar continuum (2~parameters), 
and a free parameter to fit the residual background assumed to be constant 
within the image (1~parameter), 
each image is fitted with a total of 27~parameters. 

Finally, after obtaining the best fit to each individual 2D image, we calculated 
a model image for each line using the best fit parameters, but changing the brightness parameter of the measured 
line to zero. 
This model image is subtracted from the data image, providing the best estimate of the line image in
which the contribution of all other lines and of the continuum has been removed. 
We measured the brightness in the line image using aperture photometry.
The error bars were calculated using the 2D image of the 
propagated errors provided by the instrument pipeline; these errors are
mainly dominated by the photon noise.  
The resulting brightnesses were used to obtain the transit light curves plotted and discussed 
in Sect.~\ref{Transit light curves}.

\section{Transit light curves}
\label{Transit light curves}
 \subsection{H\,{\sc i} Lyman-$\alpha$ light curve}
 \label{Lyman-alpha light curve}

\begin{figure}[tb]
\psfig{file=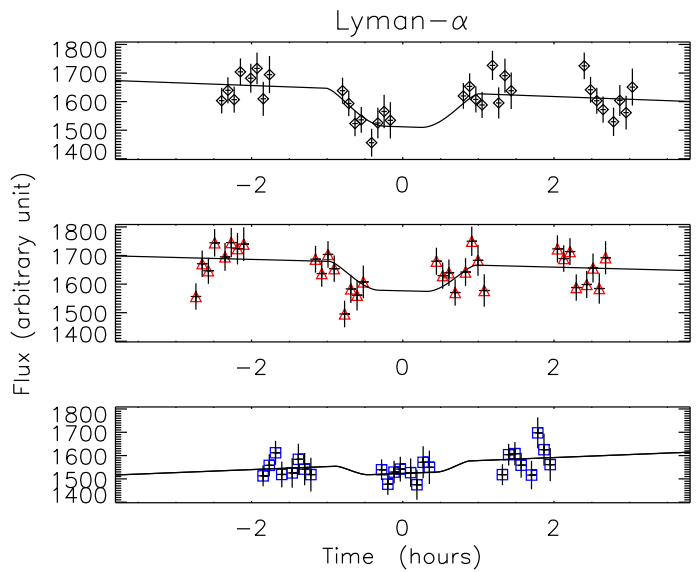,width=\columnwidth,angle=90} 
\caption[]{Lyman-$\alpha$ flux as a function of time during three transits of HD\,189733b. Different colors and symbols are for different epochs. The measurements in black diamonds, red triangles, and blue squares are for the 1$^{\rm st}$, 2$^{\rm nd}$, and 3$^{\rm rd}$ transits, respectively. The curve shows the best fit to the data assuming a linear baseline and an optically-thick disk transiting the star. The transit depth (defined by the size of the occulting disk) is free to vary from one transit to another. 
\label{Lya_reb1_pervisit}}
\end{figure}

\begin{figure}[tb]
\psfig{file=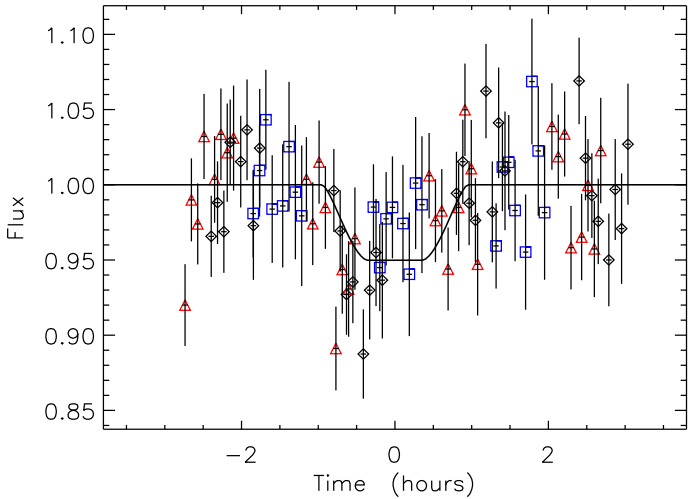,width=\columnwidth,angle=90} 
\caption[]{Flux of the Lyman-$\alpha$ line as a function of time during transits of HD\,189733b.
The flux is normalized by the linear baseline as shown in Fig.~\ref{Lya_reb1_pervisit}. The curve shows the best fit to the data by a single light curve obtained assuming the same transit depth for all transits. 
Symbols are the same as in Fig.~\ref{Lya_reb1_pervisit}.
\label{Lya_reb1}}
\end{figure}

\begin{figure}[tb]
\psfig{file=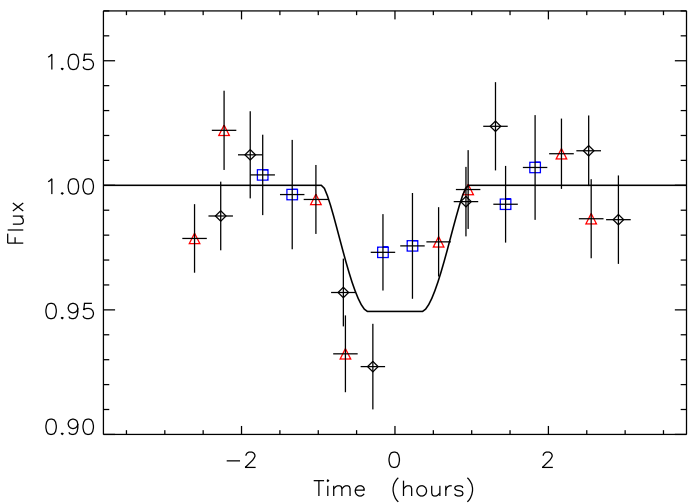,width=\columnwidth,angle=90} 
\caption[]{Same as previous plot in which the data have been rebinned by four. The possible difference of transit depth between different epochs are visible here. 
\label{Lya_reb4}}
\end{figure}

\begin{table*}[tbh]
\begin{tabular}{cccccc}
\hline
\hline
\noalign{\smallskip}
Data set & $\chi^2$ & Degree of & Absorption  & 
 $\sqrt{\Delta\chi^2}$  & $\sqrt{\Delta \chi^2}$  \\
          &         &  Freedom   & (\%)  & from 0\% abs.  & from 2.4\% abs. \\
\noalign{\smallskip}
\hline
\noalign{\smallskip}

Transit \#1 & 29.4 & 29 & 7.6 $\pm$ 1.5 & 5.7 $\sigma$ & 3.8 $\sigma$\\
Transit \#2 & 48.8 & 29 & 5.6 $\pm$ 1.7 & 3.3 $\sigma$ & 2.0 $\sigma$\\
Transit \#3 & 12.7 & 21 & 2.7 $\pm$ 1.2 & 2.2 $\sigma$ & 0.2 $\sigma$\\
  
\noalign{\smallskip}
\hline
\noalign{\smallskip}
All & 97.9 & 81 & 5.05 $\pm$ 0.75 & 6.4 $\sigma$ & 3.5 $\sigma$\\
\hline
\hline
\end{tabular}
\caption{Results from the Lyman-$\alpha$ light curve.}
\label{Table Lya}
\end{table*}

The main objective of these observations is to obtain the transit light curve 
in H\,{\sc i} Lyman-$\alpha$ and to search for an atmospheric evaporation signature. 
For each of the three transits, we plotted the resulting light curves 
(Fig.~\ref{Lya_reb1_pervisit}). We fitted each light curve 
as a function of time by a transit model defined by a linear baseline 
for the stellar Lyman-$\alpha$ flux and the occultation 
by an optically-thick disk with the planetary ephemerides. 
The results from these three independent fits are given in Table~\ref{Table Lya}.

We see that the second transit gives noisier measurements than the first one, 
and the observations were made only when the planet 
was off transit or during the partial occultation phase. 
In this transit~\#2, no measurements during the total phase of 
the planetary transit were obtained because of the Earth 
occultation of the target as seen by HST. 
In the first and third transits, the data covered all phases 
of the transit light curve, and measurements well before
and after the transit allowed for estimates of the light curve baseline. 
Measurements obtained during the total phase of the planetary transit 
are well-fitted by a transit light curve 
of an optically-thick occulting disk (Fig.~\ref{Lya_reb1_pervisit}), 
providing an estimate of the size of the H\,{\sc i} occulting cloud.

The full set of Lyman-$\alpha$ measurements can be 
phase-folded to provide a complete transit 
light curve (Figs.~\ref{Lya_reb1} and~\ref{Lya_reb4}). 
The dispersion of the data rebinned by four is roughly divided by two,
so there is no sign of red-noise in the data.  
The fit of this light curve yields a Lyman-$\alpha$ transit
depth of 5.05$\pm$0.75\%. The error bars are estimated using the $\Delta \chi^2$ method
(see e.g., H\'ebrard et al.\ 2002), and using the same method 
we can estimate the significance of the detection by calculating 
the difference of $\chi^2$ between the best fit and a fit
with no absorption at all (5$^{\rm th}$ column of Table~\ref{Table Lya}) or a fit 
with an occultation by the planetary disk alone (6$^{\rm th}$ column of Table~\ref{Table Lya}). 
Our result shows a transit depth that appears to be deeper 
than the occultation by the planetary disk alone ($\sim$2.4\%). 
Using the statistical error bars, the excess absorption is at 3.5$\sigma$ level. 
Of course, an unknown systematic effect mimicking a transit light curve with an excess
absorption cannot be excluded; nonetheless, the most likely explanation is the presence 
of a huge cloud of atomic hydrogen surrounding HD\,189733b.

The transit depths measured on individual light curves
of the transits~\#1 and~\#3 are different by about 2.5$\sigma$. This difference
is clearly visible in the plot with data rebinned by four (Fig.~\ref{Lya_reb4}). 

\subsection{Light curves in other lines}
\label{other lines}

\begin{figure}[tb]
\psfig{file=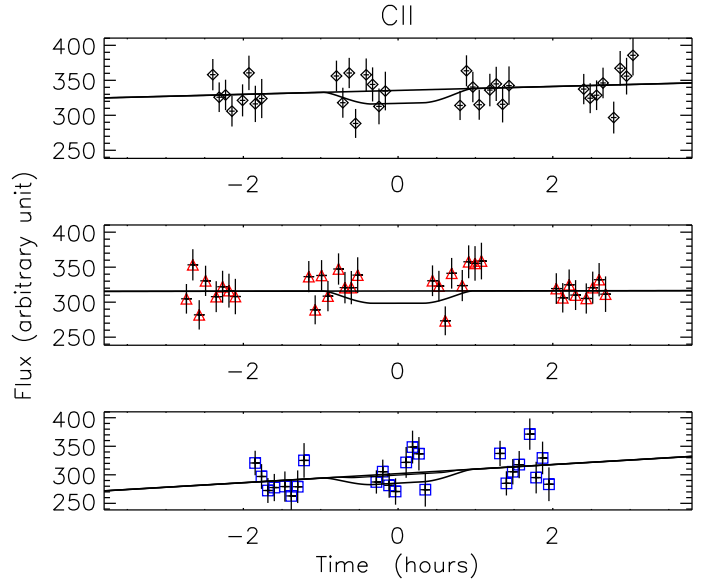,width=\columnwidth,angle=90} 
\caption[]{Plot of the flux in the C\,{\sc ii} line for the three HST transits. The data are photon-noise limited and the transit of the planet is not detected. 
Only an upper limit on the transit depth can be obtained. 
The solid line shows the transit light curve with the 3-$\sigma$ upper limit on the transit depth. 
\label{CII}}
\end{figure}

\begin{table*}[tb]
\begin{tabular}{cccccccc}
\hline
\hline
\noalign{\smallskip}
Line & $\lambda$ & N        & $\chi^2$ & Degree of & Absorption  & $\sqrt{\Delta \chi^2}$ & 3-$\sigma$ upper limit  \\
     & (\AA) & measurements &          &  Freedom  & (\%) &  from 2.4\% abs.   & (\%)  \\
\noalign{\smallskip}
\hline
\noalign{\smallskip}

O{\sc i}    & 1305  &  87  &  80.5  & 80 &  3.7 $\pm$ 2.4  &  0.5 & 11.1 \\
C{\sc ii}   & 1335  &  88  &  80.7  & 81 &  1.0 $\pm$ 1.0  &  1.3 &  5.5  \\
S{\sc iv}   & 1400  &  86  &  79.6  & 79 &  1.6 $\pm$ 1.6  &  0.8 & 11.2 \\
C{\sc iv}   & 1550  &  86  &  94.1  & 79 &  1.7 $\pm$ 1.7  &  0.5 &  7.0 \\
C{\sc i}    & 1660  &  84  &  66.2  & 77 &  0.7 $\pm$ 0.7  &  1.5 &  6.2 \\

\noalign{\smallskip}
\hline
\hline
\end{tabular}
\caption{Results from light curves of emission lines other than H\,{\sc i} Lyman-$\alpha$}
\label{Table other lines}
\end{table*}

Using the same procedure as for Lyman-$\alpha$, we measured the brightness of other stellar
emission lines and fit the light curves to measure the transit occultation depths 
(Table~\ref{Table other lines}). In these light curves, few measurements are obviously
outliers, in particular in the curve of the C\,{\sc i} line.
These outliers have been removed by 
flagging the points beyond 3$\sigma$ of the best fit. The number of points used for the
fit is given in the third column of Table~\ref{Table other lines}.

We detect a transit signature in none of these lines. 
Only upper limit on the transit depths can be derived. 
For each line, the 3-$\sigma$ upper limit corresponds to 
the transit depths that yield a $\chi^2$ of 9 above the $\chi^2$ of the best fit to the data. 
For the two most important species (O\,{\sc i} and C\,{\sc ii}), we obtain 
a 3-$\sigma$ upper limit for the transit depth of about 11\% and 5\%, 
respectively (Fig.~\ref{CII}). These two species 
are abundant and light ; therefore they should be, after hydrogen, the most abundant species 
in the upper atmosphere. In contrast to the case of HD\,209458b (Vidal-Madjar et al.\ 2004),
the non-detection of oxygen and carbon in the upper atmosphere of HD\,189733b 
precludes us from concluding anything about the nature of the escape mechanism.
Note, however, that the lines from the ground level of 
O\,{\sc i} and C\,{\sc ii} are expected to be strongly 
absorbed by the interstellar medium (see Fig.~2 of Vidal-Madjar et al.\ 2004), 
so the observed stellar emission lines are mainly from 
the excited levels O\,{\sc i}*, O\,{\sc i}**, and C\,{\sc ii}*.
Therefore, the upper limits quoted above should 
be interpreted with caution, keeping in mind that
these levels are collisionally populated in gas with typical densities 
above $\sim$$10^6$cm$^{-3}$.

\subsection{Total flux redward of Lyman-$\alpha$}

\begin{figure}[tb]
\psfig{file=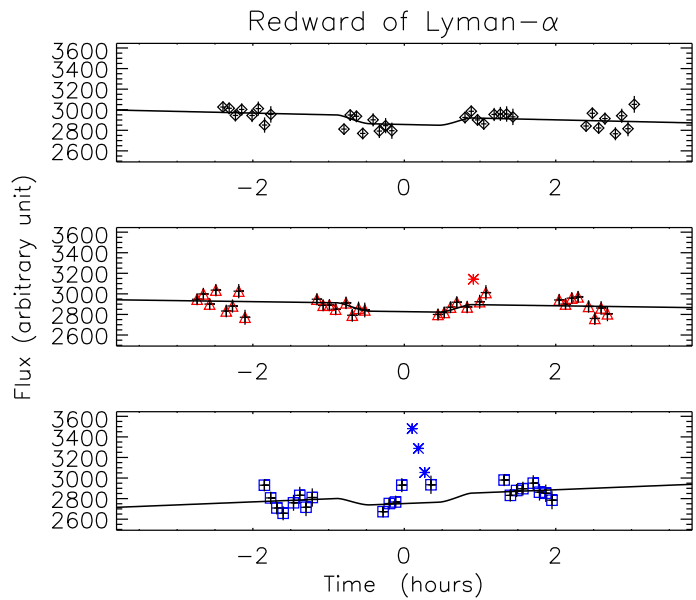,width=\columnwidth,angle=90} 
\caption[]{Plot of the flux redward of Lyman-$\alpha$ for the three HST transits. Some measurements are clearly outliers; measurements beyond 3-$\sigma$ from the best fit are plotted with stars.
Symbols are the same as in Fig.~\ref{Lya_reb1_pervisit} 
\label{redward per visit}}
\end{figure}

\begin{figure}[tb]
\psfig{file=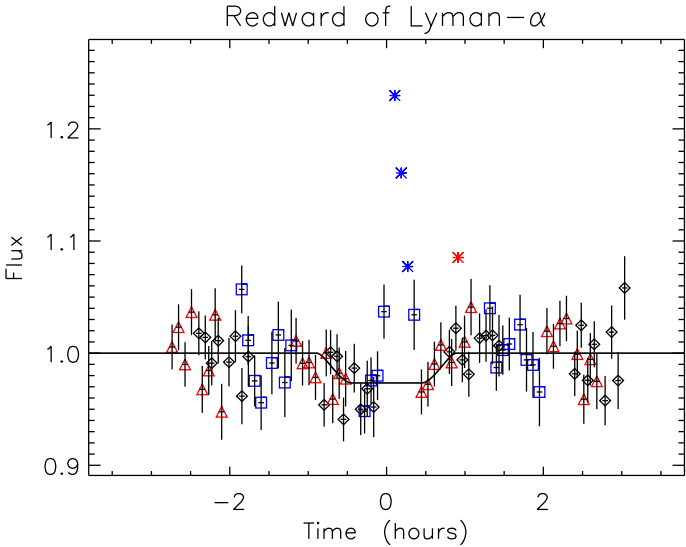,width=\columnwidth,angle=90} 
\caption[]{Plot of the flux redward of Lyman-$\alpha$. The time has been phase-folded and the flux normalized by the linear baseline shown in Fig.~\ref{redward per visit}.
Outliers at 3-$\sigma$ are plotted with stars.
\label{redward}}
\end{figure}

\begin{table*}[tbh]
\begin{tabular}{cccccccc}
\hline
\hline
\noalign{\smallskip}
Data & $\lambda$ & N      & $\chi^2$ & Degree of & Absorption  & 
 $\sqrt{\Delta\chi^2}$  & $\sqrt{\Delta \chi^2}$  \\
    & (\AA)    & measurements &         &  Freedom   & (\%)  & from 0\% abs.  & from 2.4\% abs. \\
\noalign{\smallskip}
\hline
\noalign{\smallskip}
redward of Ly-$\alpha$ & $\ge$ 1280 & 84 & 96.5 & 77 & 2.7 $\pm$ 0.6 & 4.6 $\sigma$ & 0.4 $\sigma$\\

\noalign{\smallskip}
\hline
\hline
\end{tabular}
\caption{Results from the light curve redward of Lyman-$\alpha$}
\label{Table redward}
\end{table*}

The planet transit cannot be seen in stellar emission lines 
other than Lyman-$\alpha$ (Sect.~\ref{other lines}). This is caused by the very low number of photons available in those lines. For instance, there are about 
200~photons per exposure in the O\,{\sc i} feature at 1305\,\AA ; with a dozen of exposures 
during the planetary transit, the accuracy of the transit depth is photon-noise 
limited to about $\sim$2\%.
This limitation can be overcome by the adding photons in a wide zone 
of the spectra redward of the Lyman-$\alpha$ line, including 
all these other emission lines and the stellar continuum together. 
Using the range from 1280\,\AA\ to 1700\,\AA , we find about 3000~photons 
per exposure (corresponding to about 0.5\% photon-noise limit on the transit depth 
measurement), to be compared 
with the 1700 photons per exposure in the Lyman-$\alpha$ line. 
We calculated the corresponding light curve (Figs.~\ref{redward per visit} and 
\ref{redward}) and found a clear transit signature. In this light curve, we
also see a flare-like feature superimposed on the transit light 
curve\footnote{This flare-like feature looks similar 
to a flare seen at optical wavelength 
during the transit of the planet OGLE-TR-10b (Bentley et al.\ 2009). 
This topic will be discussed further in a separate paper.}. 
This flare took place in the middle of the third visit, at the same epochs as
outliers already seen in light curves of individual lines, strengthening 
the need for flagging those measurements. 
For the spectra redward of Lyman-$\alpha$, if outliers beyond 3-$\sigma$
are flagged, we find a transit depth of 2.7$\pm$0.6\%, which is consitent with the transit of the planet 
with a depth of 2.4\% (Table~\ref{Table redward}). 
This 4.6$\sigma$-detection of the planet transit as expected 
in such data gives confidence to the detection 
of the transit seen at Lyman-$\alpha$. The calculated error bars on 
the depths of both these detected transits are also consistent 
with the theoretical limit from photon noise. 

Finally, the detection of a flare in the middle of the third transit could explain 
why the transit depth in Lyman-$\alpha$ is shallower in this last observation. If 
the measurements obtained at the epoch of the flare are flagged (the epoch 
of the 3 outliers at 3-$\sigma$ plus the two adjacent points), 
we find a Lyman-$\alpha$ transit depth of 5.7$\pm$0.9\%. With this last 
result, we would reach exactly the same conclusions.
Also, a careful inspection of individual measurements 
does not show any hint of flare in Lyman-$\alpha$ (Fig.~\ref{Lya_reb1_pervisit}) 
or in the O\,{\sc i} line, 
in contrast to carbon lines (C\,{\sc ii} plotted in Fig.~\ref{CII} and C\,{\sc iv}), 
which show suspiciously large flux at the epoch of the flare.
This behavior is similar to what has been seen for solar flares, where large
variations are seen in C\,{\sc iv} lines, while the Lyman-$\alpha$ line that originates 
in a different part of the stellar atmosphere barely shows variations 
({\it e.g.}, Brekket et al.\ 1996).  
We conclude that the flare does not affect the main results of the present paper. 
The detection of the transit of the planet in the spectra redward of Lyman-$\alpha$ 
strengthens the results at Lyman-$\alpha$.  

\section{Evaporation of HD\,189733b}
\label{Evaporation}

\subsection{The stellar Lyman-$\alpha$ emission line profile}
\label{line profile}

\begin{figure}[tb]
\psfig{file=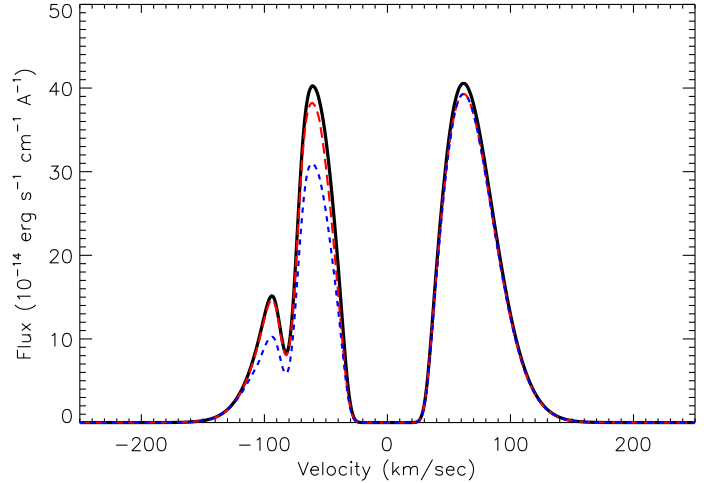,width=\columnwidth,angle=90} 
\caption[]{Plot of the theoretical spectrum of the Lyman-$\alpha$ line of HD\,189733. 
The line shape takes the absorption by the interstellar medium 
hydrogen and deuterium into account (black thick line). 
Two different spectra, including absorption by the planetary exophere 
when it passes in front of the star, are superimposed (dashed lines). 
The blue short-dashed line shows the resulting spectrum 
when the escape rate is 3$\times$10$^{8}$\,g\,s$^{-1}$ 
and the ionizing EUV flux is 0.1 times the solar value. 
With these values, in particular the low EUV flux, 
the blue side of the line is absorbed by about 25\%, 
resulting in a decrease of about 13\% in the total Lyman-$\alpha$ flux.
The red long-dashed line shows the theoretical spectrum of the Lyman-$\alpha$ line 
during transit for an escape rate of 10$^{10}$\,g\,s$^{-1}$ 
and an ionizing EUV flux 20~times the solar value. 
In this last case, the total Lyman-$\alpha$ flux decreases 
by about 5\% when the exosphere passes in front of the star. 
\label{Simulated spectrum}}
\end{figure}

As shown in Sect.~\ref{Lyman-alpha light curve}, we detect a transit signature 
in the light curve of the H\,{\sc i} Lyman-$\alpha$ stellar emission line
with a depth of $\sim$5\%, significantly deeper than the transit 2.4\% depth due
to the planetary disk alone, as seen at optical wavelengths. 
Moreover, the HD\,189733b Lyman-$\alpha$ line is a broad emission line,
with a width of $\sim$$\pm$150\,km\,s$^{-1}$. 
From the experience with HD\,209458b, we know that to interpret 
this measurement we must consider the possibility 
that the hydrogen cloud surrounding HD\,189733b only absorbs 
a fraction of this stellar line width. 
Therefore the measured 5\% absorption 
depth in the total flux of the line is possibly caused by a larger absorption of 
only a fraction of the line width (see discussion in Vidal-Madjar et al.\ 2008).
To interpret the measured absorption depth, it is therefore important 
to have an estimate of the emission line width and shape. 

The profile of the Lyman-$\alpha$ stellar emission line seen from Earth 
is composed of the profile of the intrinsic stellar line combined with 
the absorption by the interstellar atomic hydrogen and deuterium.
Interstellar hydrogen produced a narrow absorption at the line center.
Deuterium absorption is weaker and blueshifted by 80\,km\,s$^{-1}$. 
The column density of atomic hydrogen toward HD\,189733b is not known, 
but a simple relation between the stars' distance and the H\,{\sc i} column density 
can be used. Using the results of Wood et al. (2005) and a distance of 19.3~pc 
for HD\,189733b, we infer an H\,{\sc i} column of density, $N_{\rm HI}$, of about 
$10^{18.3}$\,cm$^{-2}$. 
Our conclusions are not sensitive to this assumption 
because the column density for a star at $\sim$20~parsecs 
is typically in the flat part of the curve of growth.  
The core of the stellar line is fully absorbed by the cold interstellar medium.
The width of the stellar Lyman-$\alpha$ emission line is obtained using the 
empirical relationship between the star absolute magnitude and the 
chromospheric Lyman-$\alpha$ line FWHM (Landsman \& Simon 1993). 
This relation can be written as
$W_{\rm \AA}=1.9\times 10^{-M_{V}/12.4}$, 
where the width $W_{\rm \AA}$ is given in \AA ngstroms. 
With an absolute 
V magnitude of 6.1, we infer a line width of about 0.61\,\AA , 
which is FWMH=150\,km\,s$^{-1}$, or for a Gaussian 
$\sigma_{Ly\alpha}$=64\,km\,s$^{-1}$. 
We modeled the double peak chromospheric emission line with two Gaussians
of width $\sigma_{Ly\alpha}/2$ and separated by $\sigma_{Ly\alpha}/2$. 
Combined with the interstellar absorption, we obtained the profile of the 
HD\,189733 Lyman-$\alpha$ shown in Fig.~\ref{Simulated spectrum}.

\subsection{HD\,189733b is evaporating}

We measure a 5\% absorption depth over the broad Lyman-$\alpha$ 
line with a width of $\sim$150\,km\,s$^{-1}$, and this can 
be interpreted by two extreme scenarios. 
In the first scenario, the hydrogen cloud occults the full line width. In this case, 
the 5\% absorption corresponds to a cloud of about 1.65~Jupiter radii in size 
with gas velocities reaching over 150\,km\,s$^{-1}$. 
However, at 1.65~Jupiter radii the escape velocity from
HD\,189733b is only 49\,km\,s$^{-1}$ (the escape velocity is 59\,km\,s$^{-1}$ at
one planetary radius). In this first scenario, the hydrogen
gas detected in the Lyman-$\alpha$ 
transit light curve exceeds the escape velocity, and 
is therefore not gravitationally bound to the planet, so
the detected gas must be escaping. 

In a second scenario, the velocity of the occulting gas 
does not exceed the escape velocity, and so 
the absorption only occurs in the core of the Lyman-$\alpha$ line. 
Because only 1/6 of the observed Lyman-$\alpha$ flux comes from the core
of the line within the 49\,km\,s$^{-1}$ radial velocity limit, in this scenario 
the absorption of the core must be at least 30\% to produce a final absorption
of 5\% in the full unresolved line. 
A 30\% absorption corresponds to a cloud of about 4~Jupiter radii in size, which 
significantly exceeds the size of the Roche lobe of the planet (3.3 Jupiter radii). 
In this second scenario, the gas detected in the Lyman-$\alpha$ 
transit light curve is beyond the Roche lobe, and 
is therefore not gravitationally bound to the planet, so
the detected gas must be escaping. 

An intermediate case between these two extreme scenarios can be
considered. Even if the Roche lobe was full of gas but
not beyond its borders, 
and the gas velocity was up to the escape velocity but not above it, 
the absorption depth would only be 3.3\%. Therefore, the measured 
5\% absorption depth shows
that some gas must be either beyond the Roche lobe or at a velocity 
above the escape velocity.  
The reality is most likely in between the two extreme scenarios with both 
a fraction of gas outside the Roche lobe and a fraction of gas at high velocity. 
The present observations of a 5\% absorption of the broad 
Lyman-$\alpha$ emission line demonstrates that HD\,189733b must be evaporating. 

\subsection{Numerical simulation}
\label{Numerical simulation}

To interpret the H\,{\sc i} light curve in more detail, 
we modeled the atmospheric gas escaping HD\,189733b with a numerical 
simulation including the dynamics of the hydrogen atoms. 
In this N-body simulation, hydrogen atoms are released from HD\,189733b's upper
atmosphere
with a random initial velocity corresponding to a 10\,000\,K exobase. 
This temperature corresponds to the temperature expected in the upper atmosphere
(e.g., Lecavelier des Etangs et al.\ 2004; Stone \& Proga et al.\ 2009) 
and has been actually measured for HD\,209458b (Ballester et al.\ 2007). 
In any case,
our results do not depend on this assumption because atoms are rapidly
accelerated by the radiation pressure to velocities several times higher
than their initial velocities. 
In this model, the gravity of both the star and the planet are taken into account. 
The radiation pressure from the stellar Lyman-$\alpha$ emission line is calculated
as a function of the radial velocity of the atoms. 
The self-extinction within the cloud is also taken into account.

The hydrogen atoms are supposed to be ionized by the stellar extreme-UV (EUV). 
In the simulation, their lifetime is calculated as a function of the EUV
flux, given as an input parameter to the model and expressed in solar units. 
The second input parameter is the atomic hydrogen escape rate, expressed in 
grams per second.

The dynamical model provides a steady state distribution of positions and velocities 
of escaping hydrogen atoms in the cloud surrounding HD\,189733b. From this information
we calculated the corresponding absorption over the stellar line and the corresponding
transit light curve of the total Lyman-$\alpha$ flux to be directly compared with the 
observations.  

\subsection{Results}
\label{Results}

\begin{figure}[tb]
\psfig{file=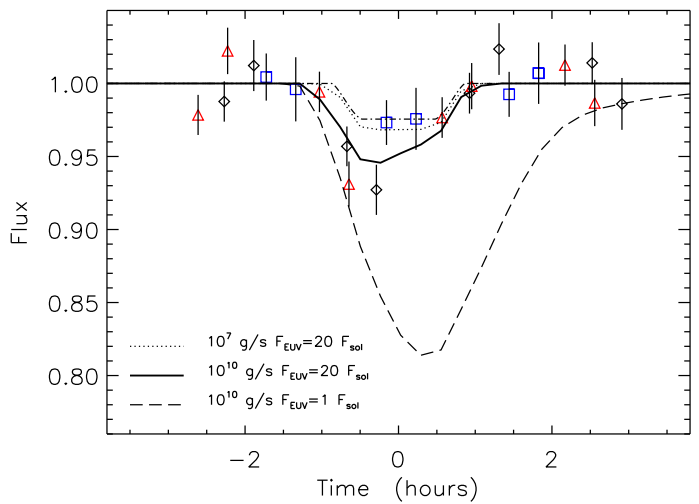,width=\columnwidth,angle=90} 
\caption[]{Plot of the theoretical light curve of the total Lyman-$\alpha$ flux 
for various escape rates and ionizing EUV fluxes. The light curve obtained 
with an escape rate of 10$^{10}$\,g\,s$^{-1}$ and an ionizing EUV flux 
of 20 times the solar value (thick line) best fits the data 
(symbols as in Fig.~\ref{Lya_reb4}) with a resulting $\chi^2$ of 91 
for 80~degrees of freedom. The dot-dashed line shows the light curve 
for a planet radius of 0.1564 times the stellar radius 
and no additional absorption by the planetary atmosphere.
\label{Fit_Model}}
\end{figure}

\begin{figure}[tb]
\psfig{file=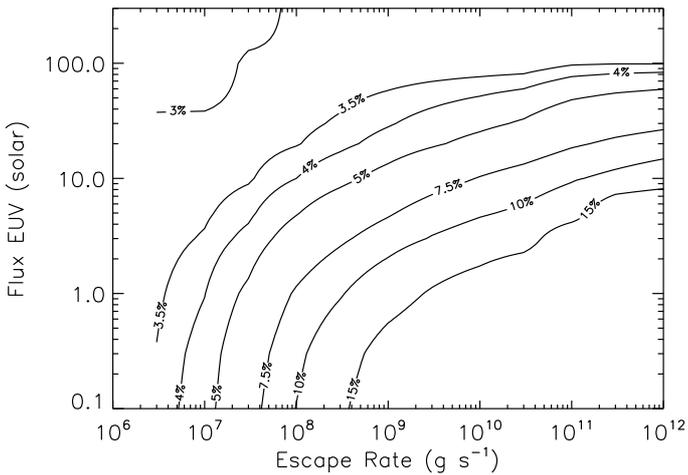,width=\columnwidth,angle=90} 
\caption[]{Plot of the absorption depth on the theoretical light curve of the stellar H\,{\sc i} Lyman-$\alpha$ line as a function of the escape rate and EUV ionizing flux.   
\label{AD_FEUV_Esc}}
\end{figure}

\begin{figure}[tb]
\psfig{file=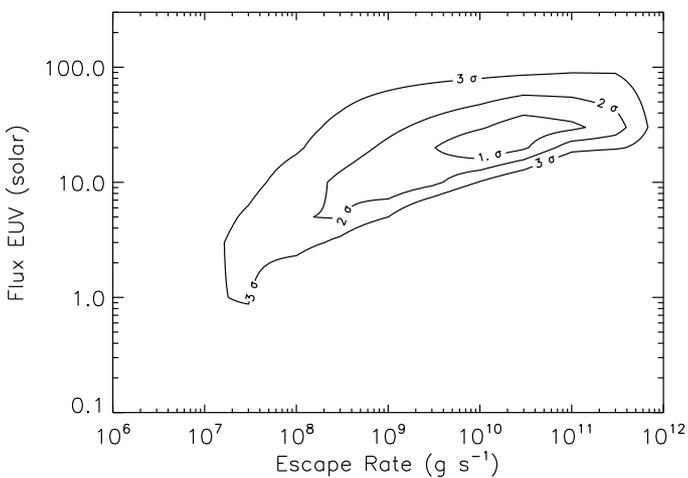,width=\columnwidth,angle=90} 
\caption[]{Plot of error bars for the estimated escape rate and EUV ionizing flux. 
The best fit is obtained for 
an escape rate of 10$^{10}$\,g\,s$^{-1}$ and an ionizing EUV flux of 20 times solar. 
To fit the data at $2\sigma$ confidence level, 
the escape is estimated to be between 10$^{8}$ 
and 3$\times$10$^{11}$\,g\,s$^{-1}$ and the EUV flux 
between 5 and 60~times the solar value. The escape rate
needed to fit the data is correlated with the ionizing flux, 
with large escape rate for a larger ionizing flux.
\label{Chi2_FEUV_Esc}}
\end{figure}

We calculated a grid of theoretical light curves as a function 
of the two input parameters, namely the escape rate and 
the EUV ionizing flux, 
and compared them to the measured Lyman-$\alpha$ flux.
The EUV flux from HD\,189733 has never been measured and is unknown. However, 
for a K-type star, it can be estimated to be 
in the range 1 to 100~times the solar EUV flux (e.g., Schmitt \& Liefke 2004). 

Using the theoretical light curves, we fitted the observational data 
with the escape rate and EUV flux 
as free parameters. For an escape rate below 10$^7$\,g\,s$^{-1}$, the occultation
by the HI cloud is negligible and the resulting Lyman-$\alpha$ 
light curve is similar to the planet occultation light curve with
a transit depth of $\sim$2.4\% (Fig.~\ref{Fit_Model}).
For a large escape rate ($\mathrm{d}M/\mathrm{d}t\ga 10^9$\,g\,s$^{-1}$) and low EUV flux
($\la$3 times solar), the planet has a long trailing H\,{\sc i}
cloud with a cometary shape, and the occultation is deeper and lasts 
longer than the optical occultation by the planetary disk alone. 
With these last conditions, the Lyman-$\alpha$ transit 
depth exceeds 10\% (Fig.~\ref{AD_FEUV_Esc}). 
Both the transit depth and duration constrains
the two parameters. Therefore,  
the fit to the Lyman-$\alpha$ light curve can be used to 
constrain both the escape rate and the ionizing flux 
(Fig.~\ref{Chi2_FEUV_Esc}). 
We find that the data require an escape rate between 
$10^9$ and $10^{11}$\,g\,s$^{-1}$ and an EUV flux 
between 10 and 40~times the solar ionizing flux (1-$\sigma$), 
with larger escape rates corresponding to larger EUV flux
(Fig.~\ref{AD_FEUV_Esc}). 
The best fit is obtained for $\mathrm{d}M/\mathrm{d}t$=$10^{10}$\,g\,s$^{-1}$ and
EUV flux $F_{\rm EUV}$=20\,$F_\odot$.
With these parameter values, the fit to the data with
the evaporating planet model yields a $\chi^2$ of 91 for 80 degrees
of freedom (Fig.~\ref{Fit_Model}). 
This fit is to be compared to the fit 
with a model of a single optically-thick 
disk for which we have a $\chi^2$ of 97.9 for 81 degrees of freedom
(Table~\ref{Table Lya}). Although the number of degrees of freedom is decreased 
by one in the evaporating planet model (because there are two free parameters,
whereas only one parameter is needed to describe the size of the
optically-thick disk), this fit is significantly better 
showing that the evaporating planet provides a 
better scenario to explain the observational data. 

\section{Discussion}

The model used to interpret the H\,{\sc i} transit light curve 
assumes a given Lyman-$\alpha$ line width 
of 0.61\,\AA\ (Sect~\ref{line profile}).
This value is rather uncertain and might be underestimated
(Landsman \& Simon 1993), and this is strengthened 
by the particularly high chromospheric activity of
this star (Henry \& Winn 2008; Boisse et al.\ 2009). 
We tested this assumption by fitting the
data as in Sects.~\ref{Numerical simulation} and~\ref{Results}, 
but using an FWMH of 1.0\,\AA\ for the simulated 
HD\,189733 Lyman-$\alpha$ stellar line. We found that, to fit the observed
light curve, a broader line requires a lower EUV ionizing flux, 
because hydrogen atoms need to reach higher velocities 
to occult the broader wings of the line before being ionized. 
We found that an EUV ionizing flux between 2 and 10 times the solar value
and similar escape rates as in Sect.~\ref{Evaporation} 
provide a good fit to the data.
Taking the uncertainty on the Lyman-$\alpha$ line width into account, 
we ended up with an EUV ionizing flux between 2 and 40 times the 
solar value (1-$\sigma$). 
This test shows that the estimate of the escape rate is not sensitive to
the assumption on the width of the stellar emission line. 

We constrained the escape rate of atomic hydrogen 
to be about $10^{10}$\,g\,s$^{-1}$, which is similar to the
escape rate previously estimated in the case of HD\,209458b.
The similarity between the two escape rates can be explained by comparing
the energy input by the stars and the physical properties of 
the planets. In effect, theoretical models of the evaporation of  
hot Jupiters agree to conclude that the escape rate can be
estimated by assuming that the EUV and X-ray energy input from the star is used by the
gas to escape the gravitation of the planet (Lecavelier des Etangs 2007). 
The potential well of HD\,189733b is about twice deeper than that of HD\,209458b
(the planet mass is 1.7 larger, and the radius is 12\% smaller). 
However, this well 
is largely compensated for by the orbital distance of HD\,189733b which 
is 1.5~times closer to its star than HD\,209458b and the
stellar type of HD\,189733 which as a K-type star is expected 
to provide a higher EUV and X-ray input energy.  
Within the uncertainties, 
these orders of magnitude show that the similarity between 
the observed evaporations of HD\,189733b and HD\,209458b 
results from their similar physical properties.  
In both cases, the observed escape rate is modest, and even if the escape of 
ionized hydrogen is not detectable in Lyman-$\alpha$, evaporation
is unlikely to change the nature of these two planets: $10^{10}$\,g\,s$^{-1}$ 
corresponds to the escape of about 0.2\% of the planets' mass in 10 billion years. 

However, the net effect of variations in EUV 
flux on the observable is not obvious. In effect, 
although an increased stellar EUV flux deposits more energy onto the
planetary upper atmosphere, and thereby increasing the heating and
consequently the atmospheric escape (e.g., Lecavelier des Etangs et al.\ 2004), 
it is also more efficient at ionizing the escaping H\,{\sc i} atoms, 
hence reducing the size of the observable H\,{\sc i} cloud 
measured during the transit. This also contributes to
explaining why we find comparable absorption depths on
HD\,189733b and HD\,209458b, although their host stars
are of different stellar types.
Here it must be recalled that the transit depth is measured on 
the Lyman-$\alpha$ line of atomic hydrogen and is interpreted in terms
of escape rate through a model in which the hydrogen ions are
produced by ionization by EUV stellar photons; therefore, 
the quoted escape rates do not include escape of hydrogen ionized 
by other mechanisms lower in the atmosphere.
   
We did not detect a transit signature of species other than hydrogen, 
because of the photon noise limit on individual lines. 
The transit signature can be seen by adding all photons redward of
the Lyman-$\alpha$ including all lines and the stellar continuum;
in this case the transit depth is consistent with the absorption 
by the planetary disk alone.
With this limitation, it is not possible to conclude 
anything about the escape mechanism, contrary to the case of HD2908458\,b 
where detection of oxygen and carbon revealed
the hydrodynamical escape (``blow-off'') of the atmosphere 
(Vidal-Madjar et al.\ 2004).

\section{Conclusion}

In summary, we have presented the 3.5$\sigma$ detection of an excess absorption 
in the Lyman-$\alpha$ transit light curve. This is interpreted in terms
of the evaporation in the upper atmosphere of HD\,189733b. 
Any detection at this level of confidence would call for an observational confirmation, which will be possible with the repaired STIS spectrograph. 
If confirmed, HD 189733b would be the second extrasolar planet 
for which evaporation has been detected.

In the present observations, the Lyman-$\alpha$ line is not resolved, 
removing any direct information on the radial velocity of the absorbing gas. 
This limitation is due to the limited spectroscopic capabilities 
of the ACS instrument in the far-ultraviolet, and the ACS was 
used because of the STIS failure. 
Fortunately, a significant improvement is expected after the refurbishment
of the Hubble Space Telescope performed last May 2009. 
With the new Cosmic Origins Spectrograph (COS) and the repaired STIS, 
new observations should allow us to address the issue of the escape mechanism 
and to probe the dynamics of escaping atoms through an improved sensitivity 
and sampling of the absorption along the Lyman-$\alpha$ line profile. 

Finally, the results from each of the three observed transits 
of HD\,189733b are not exactly the same (Table~\ref{Table Lya}). 
Since the observations were obtained at different epochs (the last transit 
was observed ten months later than the first two ones), 
this raises the question of possible time variations in the transit depth.  
From the present data, the detection of variations 
is barely significant ($\sim$2.5$\sigma$), 
and it may not be possible to draw definite conclusions
on this question. 
Nonetheless, variations would not be surprising, at least 
because of the expected variations in the EUV ionizing flux which,
for a given escape rate, significantly affects the observed transit depth (Fig.~\ref{AD_FEUV_Esc}). 
Interpretation of observations made at a single epoch is always tricky, 
and any estimate or non-detection at a given time should never be considered 
definitive. 

\begin{acknowledgements}

We warmly thank F.~Pont and F.~Bouchy for useful comments and discussions, 
and A.~Baillard and M.~Dennefeld for their OHP observations 
of the HD\,189733 field for preparing our HST observations. 

Based on observations made with the NASA/ESA Hubble Space Telescope, 
obtained at the Space Telescope Science Institute, 
which is operated by the Association of Universities for Research in Astronomy, 
Inc., under NASA contract NAS 5-26555. 
These observations are associated with program GO10869. 

D.E.\ acknowledges financial support from the French Agence Nationale 
pour la Recherche (ANR; project NT05-4\_44463) 
and from the Centre National d'\'Etudes Spatiales (CNES).
G.E.B.\ acknowledges financial support by this program through STScI grant 
HST-GO-10869.01-A to the University of Arizona.

\end{acknowledgements}

\end{document}